\documentstyle[12pt]{article}

\begin{document}

\begin{center}
{\Large 
{\bf One- and Two-proton Transfer Reactions \\
with Vibrational Nuclei\\}}
\vskip 1cm
{\Large H.D. Marta}\\
\medskip
Instituto de F\'\i sica, Facultad de Ingenier\'\i a,\\
CC 30 Montevideo, Uruguay \\
\bigskip
{\Large R. Donangelo}\\
\medskip
Instituto de F\'\i sica, Universidade Federal do Rio de Janeiro,\\
C.P. 68528, 21945-970 Rio de Janeiro, Brazil\\
\bigskip
{\Large J.O. Fern\'andez Niello, and A.J. Pacheco} \\
\medskip
Laboratorio TANDAR, Departamento de F\'\i sica,\\
Comisi\'on Nacional de Energ\'\i a At\'omica,\\
Avenida del Libertador 8250, 1429 Buenos Aires, Argentina\\
\end{center}

\vskip 1cm
\centerline{\bf ABSTRACT}
\begin{quotation}
\vskip -3mm

We extend a semiclassical model of transfer reactions to the case in
which one of the collision partners is a vibrational nucleus. The
model is applied to one- and two-proton stripping reactions in the
$^{37}$Cl + $^{98}$Mo system, for which a rapid transition from normal
to anomalous slope in the two proton transfer reaction at energies
around the Coulomb barrier is experimentally observed. This behavior
is satisfactorily reproduced by the present extension of the model.
\vskip 5mm
\noindent PACS number(s): 25.70.Hi, 24.10.Ht
\end{quotation}
\vskip 1cm
\noindent
\vfill
\medskip
\newpage

{\bf I. INTRODUCTION}
\vskip 0.2truecm

The standard treatment of subbarrier nuclear transfer
reactions~\cite{Breit} considers this process as tunneling of a
particle from the potential well created by the donor core to the one
of the acceptor core. For a given scattering angle the tunneling is
dominated by the contribution from the associated distance of closest
approach in the classical Rutherford trajectory, $D_{Ruth}$, and the
transfer probability is

\begin{equation}
P_{tr} \; \propto \; \sin (\theta /2) \; e^{-2\kappa D_{Ruth}} \; ,
\label{Ptr}
\end{equation}

\noindent
with

\begin{equation}
\kappa = \sqrt{2\mu B_{eff}/\hbar^2}
\label{kappa}
\end{equation}

\noindent
where $\mu$ and $B_{eff}$ are the reduced mass and the effective
barrier height to be traversed by the transferred particle,
respectively. When, as customary, the transfer probability is
presented as a function of $D_{Ruth}$ in a semilogarithmic plot, this
model predicts a straight line with an energy independent slope, being
the slope for two-nucleon transfer approximately twice that for
one-nucleon transfer. At large distances the experimental slopes are
generally in good agreement with the predictions of this model for
one-neutron transfer. However, deviations from the expected ratio of
two have been observed in two-nucleon transfer reactions, which in
the literature are referred to as ``slope anomalies''~\cite{Wu}.
Furthermore, some experiments show an energy dependence of the
slope~\cite{Tomasi1}.

In previous work~\cite{Marta1,Marta2,Marta3}, we were able to 
explain available proton and neutron transfer data, including the
slope anomaly, by considering the contribution of the two 
trajectories that lead to a given scattering angle (see Section II).
Thus far, we have applied this model exclusively to reactions for
which structure effects in the transfer process are assumed
unimportant.

In the present work we investigate the transfer process for the case
of vibrational nuclei, in particular one- and two- proton stripping
reactions, at energies close to the barrier. In Section II we provide
a brief review of the semiclassical theory of transfer in the case of
structureless spherical nuclei, studied before, and extend it to the
vibrational nuclei considered here. The results of the calculations
performed with the model are presented in Section III. In the last
section we draw the main conclusions and suggestions for additional
work.

\vskip 0.5truecm

{\bf II. THEORY}
\vskip 0.2truecm
{\bf a. Structureless nuclei}
\vskip 0.2truecm

In the framework of this semiclassical model~\cite{Marta1,Marta2},
the trajectories of the participant ions are determined taking into
account both the Coulomb and the nuclear part of the nucleus-nucleus
interaction. This brings, as a consequence, the possibility that more
than one trajectory contributes to a given scattering angle of the
outgoing particle. Due to absorption usually only two of them
contributes to the transfer reactions. For the nuclear optical
potential we adopt a Woods-Saxon shape with radius and strength
calculated as in Ref.~\cite{Broglia}:

\begin{equation}
R=R_p + R_t + 0.29 \, \, {\rm fm}
\label{R}
\end{equation}

\noindent
with

\begin{equation}
R_i = (1.233 A_i^{1/3} - 0.98 A_i^{-1/3}) \, \, {\rm fm}\, \, i=p,t
\label{Ri}
\end{equation}

\noindent
and

\begin{equation}
V_0 = 16 \pi \gamma \overline R a \, \, {\rm MeV}
\label{V0}
\end{equation}

\noindent
with

\begin{equation}
\gamma = 0.95 \; \left[1 - 1.8 \left({N_p-Z_p \over A_p}\right) 
\left({N_t-Z_t \over A_t}\right) \right] \,
{\rm MeV \; fm}^{-2}
\label{gamma}
\end{equation}

\noindent
and

\begin{equation}
\overline R = {R_p R_t \over R_p + R_t} \, ,
\label{Rbar}
\end{equation}

\noindent
where $a$ is the diffuseness, and $A_i$, $N_i$, and $Z_i$ are the
mass, neutron, and atomic numbers of the nucleus $i$ ($i=p$ for
projectile, $i=t$ for target), respectively.

The probability amplitude for survival from absorption due to the
imaginary part of the nucleus-nucleus optical potential, $W(r)$, is
calculated by the expression~\cite{Guidry}

\begin{equation}
a_{abs} = \exp \biggl( - {1 \over \hbar}
\int_{-\infty}^{+\infty} W(t) dt \biggr)
\label{aabs}
\end{equation}

\noindent
in which $W(t)=W(r(t))$.

We denote by $U(r)$ the potential which acts over the transferred
particle, $R_B$ the position where this potential barrier reaches its
maximum, $U_B = U(R_B)$, and $B.E.$ the binding energy of the particle
in the donor nucleus. The probability for tunneling is determined,
when $U_B + B.E. > 0$, by the WKB approximation

\begin{equation}
P_{tun}=\vert a_{tun}\vert ^2={\left( 1+ e^S \right) } ^{-1}	\label{WKB}
\end{equation}

\noindent
in which

\begin{equation}
S = 2 \int_{R_1}^{R_2} \left[ {2 \mu \over \hbar}
(U(r) + B.E.) \right] ^{1/2}  dr \; . \label {SWKB}
\end{equation}

\noindent
In the region $U_B + B.E. < 0$, the potential barrier can be
approximated by an inverted parabola, allowing us the use of the
analytic expression of Hill and Wheeler~\cite{H-W}

\begin{equation}
P_{tun} = \left[ 1+ \exp \left( 2 \pi/\hbar \omega (U_B + B.E.) \right)
\right]^{-1}	\label{HW}
\end{equation}

\noindent
with

\begin{equation}
\hbar \omega = \left( - {\hbar^2 \over \mu} \, {d^2 U(R_B) \over
{dr}^2} \right)^{1/2}.
\label{hbomega}
\end{equation}

\noindent
We have taken

\begin{equation}
U(r) = U_1(r) + U_2(D-r),	\label{U}
\end{equation}

\begin{equation}
U_j(r) = U_{C_j}(r) + U_{N_j}(r),	\label{Uj}
\end{equation}

\noindent
where the subscripts 1 and 2 refer to the donor and acceptor cores,
respectively, $D$ is the distance of closest approach between
them, and r is the spatial coordinate of the transferred particle with
respect to the donor core. $U_{C_j}$ is the Coulomb potential and
$U_{N_j}$ the nuclear potential generated by the core $j$ over the
particle. The Coulomb potential was taken as that generated by a
charged sphere of radius $1.25  A_j^{1/3}$~fm acting over a cluster
with charge $Z_{cl}$ and the nuclear part as a Saxon-Woods potential
with radius parameter $r_0 = 1.2$~fm, diffuseness $a_u = 0.63$~fm and
depths

\begin{equation}
U_{0_j} = Z_{cl} V_{0_j}(+1) + N_{cl} V_{0_j} (-1)
\label{U0j}
\end{equation}

\noindent
where

\begin{equation}
V_{0_j}(\tau_z) =
\left[ 51 - 33 \tau_z \, {N_j - Z_j \over {A_j} } \right]
\; {\rm MeV} \;.	\label{V0j}
\end{equation}

\noindent
is the nuclear potential generated by the core $j$ with mass number
$A_j$, charge $Z_j$ and neutron number $N_j = A_j - Z_j$ acting over
the neutrons ($\tau_z$ = +1) and the protons ($\tau_z$ = -1) of the
transferred cluster with mass number $A_{cl}$, charge $Z_{cl}$, and
neutron number $N_{cl} = A_{cl} - Z_{cl}$.

As detailed in Ref.~\cite{Marta1}, except in cases in which the
measurements are done with high angular resolution, the transfer
probability can be approximated by the incoherent sum of contributions
by each trajectory leading to a given scattering angle

\begin{equation}
P_{tr} ( \theta ) \approx \sum P_{tun}( \theta ) \;
\mid a_{abs}( \theta ) \mid^2 \, . 	\label{trp}
\end{equation}

\noindent
This expression is employed in the calculations presented in this
work.
\vskip 0.5truecm

{\bf b. Vibrational nuclei}
\vskip 0.2truecm

For a transfer process in which one of the participant ions is a
vibrational nucleus (the target nucleus, to fix ideas) the above
mentioned model can be extended in the following way. Assuming a
quadrupole vibrational mode, we parametrize the radius of the target
as

\begin{equation}
R_t(\alpha_{20}) = R_{t} (1 + \alpha_{20} Y_{20}(\theta) ) .
\label{Rt}
\end{equation}

\noindent
The internal Hamiltonian of the target may be written as ~\cite{Bohr}

\begin{equation}
H_{int} = {1\over 2} ( B_{2} \vert \dot \alpha _{20} \vert^2 + C_{2}
\vert \alpha _{20}\vert^2 ) .
\label{Hint}
\end{equation}

\noindent
from which the zero-point amplitude is

\begin{equation}
\alpha_{20}^0 = \left( {E_{2+}\over {2 C_2}} \right)^{1\over 2}
\label{alpha}
\end{equation}

\noindent
The target radius $R_{t}$ is given by Eq.~(\ref{Ri}), and $E_{2+}$ is
the transition energy between the first excited and the ground state.

In this kind of reaction the vibration is very slow in comparison to
the translational motion of the projectile and the tunneling process
is dominated by those trajectories for which the distance between the
surfaces of projectile and target at the point of closest approach are
smallest. Therefore the most relevant axis of vibration is that
directed from this point to the center of the target. In this spirit
we substitute the spherical harmonic in Eq.~(\ref{Rt}) by its maximun
value $\sqrt{5 / 4\pi}$.

With these prescriptions we now calculate the transfer probability as
described for structureless nuclei, by substitution of
Eqs.~(\ref{aabs}) and (\ref{WKB}) or (\ref{HW}) into Eq.~(\ref{trp}).
The radius of the target must be replaced by $R_t(\alpha_{20})$, which
implies an additional degree of freedom in the calculation. Then, we
solve the classical equations of motion using the Hamiltonian

\begin{equation}
H = T_r + H_{int} + V(r,\alpha_{20})
\label{H}
\end{equation}

\noindent
where $T_r$ is the kinetic energy of the relative motion and
$V(r,\alpha_{20}) = V_C + V_N $ includes the potential energy of the
relative motion and the interaction between this and the vibrational
mode. We approximate

\begin{equation}
V_C = Z_p Z_t e^2 \; \left({1\over r} + {3\over 5}
{R_p \; R_t(\alpha_{20})\over r^3} \right)
\label{VC}
\end{equation}

\noindent
for the Coulomb interaction and write

\begin{equation}
V_N =
{V_0 \over 1 + e^{{[r-R_p-R_t(\alpha_{20})]/a}}}
\label{VN}
\end{equation}

\noindent
for the nuclear part. In the calculation of the absorption, the
imaginary part of the nuclear optical potential acting between the
incident ions is taken as

\begin{equation}
W_N =
{W_0 \over 1 + e^{{[r-R_p-R_t(\alpha_{20})]/a}}}
\label{WN}
\end{equation}

We integrate the resulting coupled differential equations with the
initial condition $\alpha(t=-\infty)=\alpha_{20}^0 \; \cos \phi$,
where $\phi$, the initial phase, can take any value in $[0,2\pi]$ with
equal probability.
For each phase $\phi$ the distance of closest approach and the value
of $\alpha_{20}$ at this time are calculated, in order to calculate
the transfer probability for this trajectory.
For a given scattering angle the contributions are summed as in
Eq.~(\ref{trp}) and the final transfer probability is obtained by
averaging over all phases.
\vskip 0.5truecm

{\bf III. RESULTS AND DISCUSSION}

\vskip 0.2truecm

We apply this model to the vibrational nucleus $^{98}$Mo, studied by
means of the $^{37}$Cl + $^{98}$Mo reaction. Data for reactions with
Mo isotopes, including this system, at energies close to the barrier,
were recently measured by the SUNY group~\cite{Liang,Mahon}. It was
found that, in the case of two-proton stripping reactions, there is an
abrupt change in slope at approximately the energy corresponding to
the Coulomb barrier (E$_{lab}\approx 117$~MeV). A calculation as
described in Section II.a, {\it i.e.} without considering the
vibrational character of the Mo nuclei, yields a value for the energy
at which the slope changes higher than the observed one.

In Fig.~1 we show the experimental transfer probabilities divided by
sin($\theta_{c.m.}$/2) for one-proton (circles) and two-proton
(squares) stripping reactions in the $^{37}$Cl + $^{98}$Mo
system~\cite{Mahon}. Also shown are the theoretical results calculated
as described in Section II.b and normalized to the data (full lines).
We used the parameters of Temmer and Heydenburg~\cite{Temmer} for
$^{98}$Mo, $E_{2+}$ = 0.786 MeV, $C_2$ = 70 MeV. For the optical
potential we take a diffuseness $a$~=~0.7~fm, and a strength for the
imaginary part $W_0$~=~70~MeV. As shown in Fig.~1, a remarkably good
agreement between the calculated and the experimental points is
obtained at large values of $D_{Ruth}$. In this approach $D_{Ruth}$ is
only a parametrization of the scattering
angle~\cite{Marta1,Marta2,Marta3}.

We consider particularly interesting that our calculation reproduces
the abrupt change in slope observed at bombarding energies around
the Coulomb barrier.
To understand the origin of this effect we should remember that,
for a given deflection angle, there are two trajectories contributing
to the transfer probability~\cite{Marta2}.
One is essentially a Rutherford trajectory while the other feels
more strongly the nuclear optical potential.
At lower energies the dominant contribution to the transfer probability
is the Coulomb trajectory, which gives the energy independent slope of
Eq.~(1), while at higher energies the nuclear trajectory dominates.
As shown in Ref.~\cite{Marta2} there is a rapid transition between
these two regimes, which, for a structureless nucleus takes place at
an energy well above the Coulomb barrier.
In the case of a vibrational nucleus, the effective radius increases,
which decreases the effective barrier, bringing, as a consequence,
the transition in slope to lower energies.
\vskip 0.5truecm

{\bf IV. CONCLUSIONS}
\vskip 0.2truecm

An extension of a simple semiclassical model including some nuclear
structure effects, quadrupole vibrations to be specific, was proved to
be quite succesful. Applications to other degrees of freedom, such as
octupole vibrations, can be implemented in essentially the same way.
One could, in a similar spirit, include other nuclear properties, such
as deformation. The practical problem with the nuclear deformation
case is that the trajectories depend on the orientation of the
deformation axis given by two Euler angles. The averaging process
should then be performed over two parameters, the initial values of
the two Euler angles and, consequently, one would need to consider a
very large number of trajectories contributing to a given scattering
angle, thus making the problem much more difficult to treat.

\vspace*{2cm}
The authors thank Dr. Linwood Lee, from SUNY, for calling attention
to their data and furnishing detailed tables that greatly facilitated
our analysis. They also acknowledge financial support from PEDECIBA
and the Programa Cient\'\i fico-Tecnol\'ogico CONICYT-BID (Uruguay)
(H.D.M. and R.D.), CSIC (UDELAR, Uruguay) (H.D.M.), 
MCT/FINEP/CNPq(PRONEX) under contract 41.96.0886.00, and ICTP (R.D.). 
Two of us (J.O.F.N. and A.J.P.) are members of the Consejo Nacional 
de Investigaciones Cient\'\i ficas y T\'ecnicas (CONICET), Argentina.

\newpage

{\Large \bf Figure Caption}
\begin{description}

\item[ Fig. 1]  One- and two-proton transfer probability, divided by
sin($\theta_{c.m.}$/2), as a function of $D_{\rm Ruth}$ for the
$^{37}$Cl + $^{98}$Mo reaction at three laboratory energies. Symbols
represent the experimental data of Ref.~\cite{Mahon}, and lines are
the theoretical calculations described in this work, and normalized to
the data.
\end{description}

\newpage

\begin{thebibliography}{99}

\bibitem{Breit} G. Breit and M. E. Ebel, Phys. Rev. {\bf 103}, 679 (1956).

\bibitem{Wu} C. Y. Wu, W. von Oertzen, D. Cline, and M. W. Guidry,
		Ann. Rev. Nucl. Part. Sci. {\bf 40}, 285 (1990).

\bibitem{Tomasi1} D. Tomasi, J. O. Fern\'andez Niello, A. J. Pacheco,
		D. Abriola, J. E. Testoni, A. O. Macchiavelli,
		O. A. Capurro, D. E. Di Gregorio, M. di Tada,
		G. V. Mart\'{\i}, and I. Urteaga,
		Phys. Rev. C {\bf 54}, 1282 (1996).

\bibitem{Marta1} H. D. Marta, R. Donangelo, D. Tomasi,
		J. O. Fern\'andez Niello and A. J. Pacheco,
		Phys. Rev. C {\bf 54}, 3156 (1996).

\bibitem{Marta2} H. D. Marta, R. Donangelo, D. Tomasi,
		J. O. Fern\'andez Niello and A. J. Pacheco,
		Phys. Rev. C {\bf 55}, 2975 (1997).

\bibitem{Marta3} H. D. Marta, R. Donangelo, D. Tomasi,
		J. O. Fern\'andez Niello and A. J. Pacheco,
		Phys. Rev. C {\bf 58}, 601 (1998).

\bibitem{Broglia} R. A. Broglia and A. Winther, {\it Heavy Ion
		Reactions} (Addison-Wesley, Reading, MA, 1991).

\bibitem{Guidry} M.W. Guidry, R.W. Kincaid, R. Donangelo,
		Phys. Lett. {\bf B 150}, 265 (1985).

\bibitem{H-W}   D.L. Hill and J.A. Wheeler, Phys. Rev. {\bf 89},
		1102 (1953).

\bibitem{Bohr} A. Bohr and B. Mottelson, {\it Nuclear Structure}
		(Benjamin, Reading,1975), Vol. II, p. 683.
		
\bibitem{Liang} J. F. Liang, L. L. Lee, Jr., J. C. Mahon,
		and R. J. Vojtech, Phys. Rev. C {\bf 50}, 1550 (1994).

\bibitem{Mahon} J. C. Mahon, L. L. Lee, Jr., J. F. Liang, C. R. Morton,
		N. T. P. Bateman, K. Yildiz, and B. M. Young,
		J. Phys. G {\bf 23}, 1215 (1997).

\bibitem{Temmer} G. M. Temmer and N. P. Heydenburg,
		Phys. Rev. {\bf 104}, 967 (1956).

\end{thebibliography}
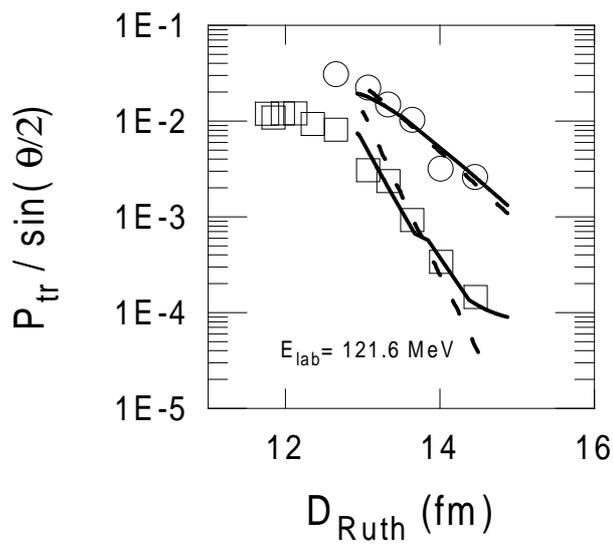
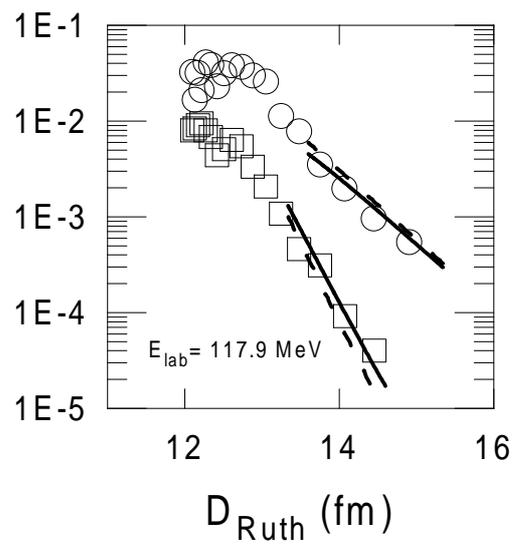
\end{document}